# The Solar Ultraviolet Imaging Telescope on-board Aditya-L1

**Durgesh Tripathi[1,*], A. N. Ramaprakash[1], Aafaque Khan[1], Avyarthana Ghosh[1,2], Subhamoy Chatterjee[3], Dipankar Banerjee[3], Pravin Chordia[1], Achim Gandorfer[4], Natalie Krivova[4], Dibyendu Nandy[2], Chaitanya Rajarshi[1] and Sami K. Solanki[4]**

[1]Inter-University Centre for Astronomy and Astrophysics (IUCAA), Post Bag-4, Ganeshkhind, Pune 411 007, India
[2]CESSI, Indian Institute of Science Education and Research, Kolkata 741 246, India
[3]Indian Institute of Astrophysics, 2nd Block, Koramangala, Bengaluru 560 034, India
[4]Max-Planck Institute for Solar System Research, Justus-von-Liebig-Weg 3, 37077, Göttingen, Germany

**The Solar Ultraviolet Imaging Telescope (SUIT) is an instrument on-board Aditya-L1 mission of ISRO that will measure and monitor the solar radiation emitted in the near ultraviolet wavelength range (200–400 nm). SUIT will simultaneously map the photosphere and chromosphere of the Sun using 11 filters sensitive to different wavelengths and covering different heights in the solar atmosphere and help us understand the processes involved in the transfer from mass and energy from one layer to the other. SUIT will also allow us to measure and monitor spatially resolved solar spectral irradiance that governs the chemistry of oxygen and ozone in the stratosphere of the Earth's atmosphere. This is central to our understanding of Sun–climate relationship.**

## Introduction

IT was Galileo who observed the Sun through a telescope for the first time over 400 years ago, leading to the discovery of sunspots. With extensive observations made from the ground dating back to the 17th century and space-based studies in the last few decades, we have collected vast amounts of data and made many fascinating discoveries about our neighbouring star. However, we only have a limited understanding of many physical phenomena associated with the energetics and dynamics of the Sun and its impact on our planet.

The atmosphere of the Sun presents us with a number of physical phenomena of great importance. One of the most important questions in solar astrophysics is regarding the existence of the extremely hot corona – the uppermost atmosphere which is above the merely 6000 K photosphere (Figure 1)[1]. Due to such high temperature, the Sun radiates at high energies such as ultraviolet (UV) and X-rays. The existence of the higher temperature of upper layers of the atmosphere above the cooler layers is one of the most outstanding problems in astrophysics since its discovery in 1940s.

It is now known that the solar atmosphere is highly dynamic and shows eruptions at various spatio-temporal scales ranging from sub-arcsec (<700 km) to the solar radius or even larger (>700,000 km). While the small-scale structures likely play important roles in transferring mass and energy from one layer to another, the large-scale eruptions could have devastating effects on space weather and geo-space climate that affect satellite communication, cause electric power blackout, etc. An important goal of solar physics is to be able to predict such large-scale eruptions from the Sun and thereby mitigate their impacts.

The Earth's atmosphere absorbs the high-energy radiation in X-rays and UV. The schematic presented in Figure 2 shows the reach of the UV radiation in the atmosphere of the Earth. The UV radiation from the Sun can be

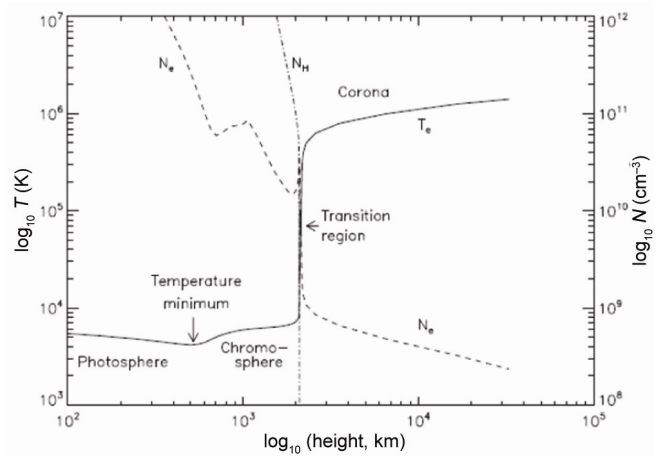

**Figure 1.** Temperature and density profile of a 1D stratified solar atmosphere as a function of height (from Aschwanden[1]).

---

divided into three regions, namely UV-A, UV-B and UV-C. As can be seen from Figure 2, the atmosphere of the Earth absorbs almost all the UV-C and UV-B radiation. The main source of absorption below 200 nm is $O_2$. For radiation between 200 and 240 nm, both $O_2$ and $O_3$ play major roles. The radiation above 242 nm is mostly absorbed by $O_3$. UV radiation above 310 nm penetrates through the Earth's atmosphere. Therefore, the radiation from the Sun within the wavelength range 200–400 nm is central in order to understand the effects of solar radiation on the dynamics and chemistry of the terrestrial atmosohere as well as the climate dynamics on the Earth.

Total energy output from the Sun at wavelengths below 400 nm is just about 8% of the total solar energy output, i.e. total solar irradiance (TSI). However, more than 60% variability is recorded in the radiation below 400 nm over a solar cycle[2]. Note that the variability in TSI over a solar cycle is about 0.1%. Being a critical input to climate models, the magnitude of the irradiance variability in the UV-B and UV-C is still a matter of debate. Measurements by the SIM instrument on SORCE suggested changes between 2007 and 2004 which were a factor of 3–6 stronger than possibly predicted by state-of-the-art models[3]. We note here that these measurements are based on Sun-as-a-star observations, i.e. considering the Sun as a point source and without resolving the individual surface structures that cause the irradiance changes. Modern-day images of the Sun show that the solar atmosphere consists of a menagerie of structures with different temperatures, densities and length scales, and they radiate at different temperatures. Therefore, the results obtained using Sun-as-a-star measurements, though of extreme importance, do not really provide insights into the causes of the observed variability.

The Solar Ultraviolet Imaging Telescope (SUIT)[4], on-board the Aditya-L1 mission is aimed to study the radiation emitted by the Sun in the wavelength range 200–400 nm (Figure 3). The telescope is being developed at the Inter-University Centre for Astronomy and Astrophysics (IUCAA), Pune in collaboration with the Centre of Excellence in Space Sciences (CESSI), Indian Institute of Science Education and Research (IISER), Kolkata; Indian Institute of Astrophysics (IIA), Bengaluru and various agencies of the Indian Space Research Organisation (ISRO).

SUIT will provide full-disk observations of the Sun in the near ultraviolet (NUV) wavelength range 200–400 nm in 11 wavelength passbands (Table 1). The SUIT instrument will open up a new observational window for solar observations at NUV wavelengths, without being subject to attenuation due to the Earth's atmosphere. SUIT will have a combination of medium and narrow band filters that cater to different scientific objectives. It aims at providing near-simultaneous full-disk images of the lower and middle layers of the solar atmosphere, namely photosphere, chromosphere and lower transition region.

The specific science goals to be addressed by SUIT are as follows: (a) Coupling and dynamics of the solar atmosphere: What are the processes through which the energy is channellized and transferred from the photosphere to the chromosphere and then to the corona? (b) Prominence studies from SUIT: What are the mechanisms responsible for stability, dynamics and eruption of solar prominences? (c) Initiation of CMEs and space weather: What is the kinematics of erupting prominences during the early phase. (d) Sun–climate studies with SUIT: How strongly does the solar spectral irradiance of relevance for the Earth's climate vary?

The SUIT instrument is designed to provide a spatial resolution of about ~1.4 arcsec. This will enable us to resolve structures of about 1000 km in size with a signal-to-noise ratio of about 100 : 1.

SUIT has two main sub-units, namely the optical bench and payload electronics. The optical bench will be mounted on the spacecraft deck along with other payloads.

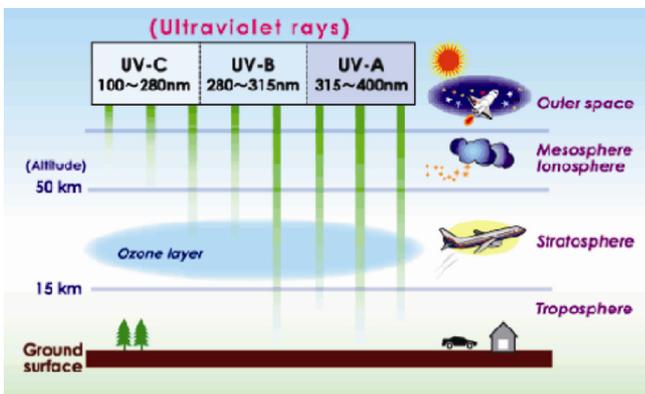

**Figure 2.** UV rays and their reach into the atmosphere of the Earth. Credit: Centre for Global Environment Research, National Institute for Environmental Studies, Japan.

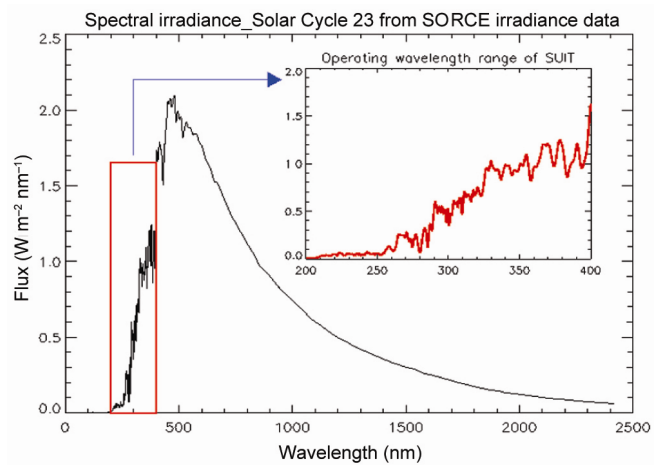

**Figure 3.** Solar radiation spectrum from 0 to 2500 A. The red box shows the wavelength that will be observed by Solar Ultraviolet Imaging Telescope (SUIT).

Table 1. Filter passbands that will be used by SUIT to map different atmospheric layers of the Sun and the corresponding science goals

| Spectral channels (nm) | Bandpass (nm) | Science |
|---|---|---|
| 214 Upper photosphere | 5 | Dynamics of the magnetic bright points in the photosphere |
| 274.7 Blue wing of Mg lines | 0.4 | Chromospheric and lower transition region dynamics, waves, shocks, filaments and prominences |
| 279.6 Mg II h line | 0.4 | |
| 280.3 Mg II k line | 0.4 | |
| 283.2 Red wing of the Mg II lines | 0.4 | Sun–earth climate connection |
| 300 Sunspots | 1 | Sunspot |
| 388 $T_{min}$ | 1 | Monitoring the magnetic flux proxies |
| 396.85 Ca II line | 0.1 | Chromosphere |
| 200–242 $O_2$ Herzberg Continuum $O_3$ Hartley band | 42 | Sun–Earth climate connection: photodissociation of oxygen and ozone in the stratosphere |
| 242–300 $O_3$ Hartley band | 58 | |
| 320–360 $O_3$ Hartley–Huggins bands | 40 | |

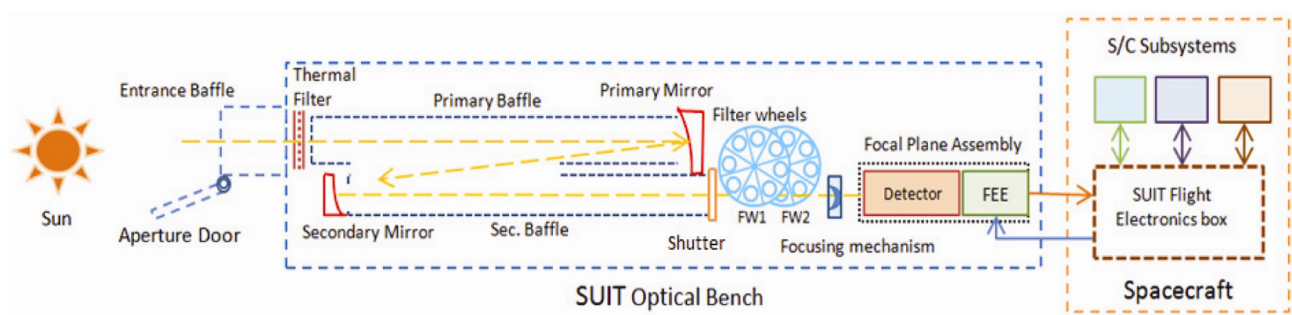

**Figure 4.** Functional diagram of the SUIT.

It is a two-mirror off-axis telescope that is designed to observe the Sun in the required wavelength range, at the demanded spatial resolution, using a passively cooled charge-couple device (CCD) detector. The key components of the telescope are entrance-door mechanism, thermal filter, primary and secondary mirrors, shutter mechanism, baffles, two-filter wheel assemblies, focusing mechanism and the focal plane assembly.

As shown in Figures 4 and 5, solar radiation enters the payload from the aperture to reach the thermal filter. As mentioned earlier, the total energy radiated in the wavelength range below 400 nm is just about 8% of the total energy radiated from the Sun. If all the radiation from the Sun is allowed to enter the optical cavity, the mirrors and the detector would get damaged due to overheating. The metal–dielectric thermal filter is designed to reflect most of the solar flux below 200 nm and above 400 nm. In addition, it will also cut down fluxes between 200 and 400 nm and only 1% of the flux in this region will be transmitted to the main optical chamber of SUIT.

There are two filter wheels to accommodate 11 science filters (Table 1), four neutral density filters and an opaque block. The neutral density filters are required to balance the fluxes at the detector in different passbands. This is necessary as the solar flux increases by a factor of 20 from 200 nm to 400 nm (Figure 3). The two filter wheels can be driven independently to achieve the desired combination of a science filter with a neutral density filter. The shutter mechanism is used to control the exposure with different filter combinations to achieve the desired signal-to-noise ratio.

The telescope is being designed to give high-resolution images of the Sun. However, due to variability in the thermal environment the telescope could get defocused, leading to loss in optical performance of the payload. The focusing mechanism, which consists of a single lens mounted on a linear stage, has been designed to compensate for defocus due to variation in the equilibrium temperature of the payload. It can also be used to compensate for any defocus that may be caused by misalignments due to launch vibrations.

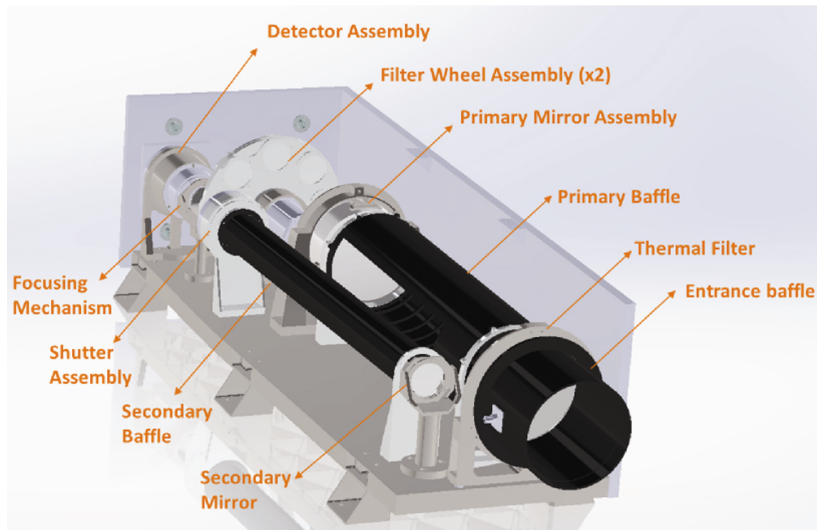

**Figure 5.** SUIT payload with all the subsystems.

The transmitted flux from the thermal filter passes through the optical system and eventually reaches the CCD detector which will be maintained at –50°C by a passive thermal control system. The CCD detector will be readout by processing electronics that will be located below the top deck of the spacecraft. The front-end electronics present in the vicinity of the CCD detector will be interfaced with the readout electronics through interface cables for data and power. The processing electronics will also control the mechanisms, execute the operational sequence of the payload and relay the data to the storage on-board the spacecraft.

The payload will be operated by the processing electronics according to predefined sequences and operational modes. For each exposure, the two filter wheel mechanisms will independently move a desired combination of science and neutral density filters into the beam path. While the filter wheels are being moved, the beam will be blocked by the shutter mechanism. Once the desired filters are in position, the shutter will open for a preprogrammed duration to expose the CCD detector. After exposure, the shutter will remain closed while the detector is read and the filter wheels are moved into the position for the next exposure according to the operational mode.

SUIT is being designed to observe the Sun 24 × 7. Every 30 min, SUIT will provide full-disk images of the Sun in all its 11 filters. These will be useful for long-term study of spatially resolved solar spectral irradiance. In addition, at every 30 s or so, region-of-interest images will be taken in eight narrow-band filters to study the dynamics of the solar atmosphere. Moreover, there will be modes of observations that will be driven by specific science proposals. In order to observe flares, an on-board intelligence is being developed to automatically locate the flares on the surface of the Sun. Under the optimized conditions, SUIT will produce about 40 GB of data in 24 h, which is larger than the share allotted to it from the mission. Therefore, an on-board data compression technique is employed.

Spatially resolved full-disk images of the Sun in 200–400 nm wavelength range have never been obtained previously. The observations recorded by SUIT will allow studies of spatially resolved solar spectral irradiance in the wavelength range 200–400 nm. On the one hand, this is essential for understanding the chemistry of oxygen and ozone in the stratosphere of the Earth, and the basis of Sun–climate forcing. On the other hand, these solar observations are crucial for exploring energy transfer, coupling and dynamics through the solar photospheric and chromospheric connection.

ACKNOWLEDGEMENTS. This work was carried out at IUCAA, Pune under the umbrella of the Max-Planck Partner Group of MPS. We thank ISRO for providing the Aditya-L1 mission opportunity and funding for the development of the SUIT payload. CESSI is funded by the Ministry of Human Resource Development, Government of India.

doi: 10.18520/cs/v113/i04/616-619